\documentclass[12pt]{iopart}
\usepackage{graphicx,amssymb} 

\begin{document}

\title{Correcting the polarization effect in low frequency Dielectric Spectroscopy}

\author{Camelia Prodan$^1$ and Corina Bot$^1$}

\address{Physics Department, New Jersey Institute of Technology, Newark, NJ 07102}

\begin{abstract}
We demonstrate a simple and robust methodology for measuring and analyzing the polarization impedance appearing at interface between electrodes and ionic solutions, in the frequency range from 1 to $10^6$ Hz. The method assumes no particular behavior of the electrode polarization impedance and it only makes use of the fact that the polarization effect dies out with frequency. The method allows a direct and un-biased measurement of the polarization impedance, whose behavior with the applied voltages and ionic concentration is methodically investigated. Furthermore, based on the previous findings, we propose a protocol for correcting the polarization effect in low frequency Dielectric Spectroscopy measurements of colloids. This could potentially lead to the quantitative resolution of the $\alpha$-dispersion regime of live cells in suspension.
\end{abstract}


\maketitle

\section{Introduction}
 
	Dielectric  Spectroscopy (DS) is an experimental technique that records the low frequency variations of the complex dielectric permittivity of a sample as function of frequency, which we call the DS curves. Because it is noninvasive, DS is widely used in many areas, including biophysics, pharmacology and geophysics [1-8]. The low frequency DS measurements are particularly interesting because most of the colloidal suspensions exhibit here an $\alpha$-relaxation. The shape of the DS curves near the $\alpha$-relaxation depends, for example, on the material, shape and size of the colloidal particles. For live cells suspensions, the DS curves depend strongly on the membrane and several other important cell parameters [9-11].
	
Unfortunately, the quantitative interpretation of the DS measurements is limited by the inherent contamination from electrode polarization, which is the apparition of a polarization impedance at the interface between the electrode and electrolyte. While this effect is present at all frequencies, it become stronger as the frequency is lowered. To obtain the intrinsic DS curves of a samples, the polarization effect has to be accounted for and removed from the data. This problem is of great interest to physicists from many areas of research, as indicated by the large amount of research published on this subject [14-19]. If this can be accomplished, the experimental DS data can be quantitatively analyzed with theoretical models [9-13] and such combination could resolve absolute values (as opposed to relative values) for several parameters of great interest to scientists working with colloidal suspensions, including live biological cells. The present study is part of our effort of resolving the $\alpha$-region of DS curves for live cell suspensions, where most of the available experimental methods encounter difficulties (see the strong critiques of Ref.~[20]).
	
The present paper proposes a technique to measure, analyze and remove the electrode polarization error at low frequencies. We report applications of the proposed technique to samples with known dielectric behavior such as saline solutions and buffers. Using these data, we map the dependence of the polarization impedance on different parameters such as frequency, the applied voltage and the electrode separation distance. We also demonstrate the robustness of the technique. Pure water and water with low and high concentrations of KCl are measured in order to study the dependence of the polarization error on the electrolyte's conductivity. As we shall see in this paper, the polarization impedance is inversely proportional to the conductivity of salty water. We also measure HEPES, a widely used buffer in life sciences. This is particularly interesting because HEPES have the same conductivity as some of our KCl solutions but different molecular composition.  As we shall see, the polarization error has a different behavior for this case. We also propose a protocol for measuring, analyzing and removing the electrode polarization for colloidal suspensions. We report an application to E. coli cells in aqueous suspension. 

Let us end this section by briefly going through the existing methodologies for polarization removal and their successful applications to different systems. A detailed presentation of the existing electrode polarization removal methods can be found in the reviews of Refs.~[21-23]. Several methodologies are based on the observation that, in the very low frequency range, the polarization impedance $Z_{\mbox{\tiny{p}}}$ behaves as a frequency dependent RC circuit, with the resistor $R_{\mbox{\tiny{p}}}$ and capacitor $C_{\mbox{\tiny{p}}}$ in series and having the frequency dependent values: $C_{\mbox{\tiny{p}}}$=$A\omega^{-m}$ , $R_{\mbox{\tiny{p}}}$=$B\omega^{-n}$ [24]. The parameters A, B, $m$, $n$ depend on the electrode and electrolyte properties, as well as on their interaction. Since the polarization impedance is physically localized at the interface between the electrodes and the sample, $Z_{\mbox{\tiny{p}}}$ can be set in series with the sampleÕs impedance $Z_{\mbox{\tiny{s}}}$, to give the total measured impedance Z. In this context then, to remove the polarization errors, one has to develop methodologies for determining the parameters A, B, m and n that characterize the polarization impedance. A fitting methodology was designed and tested in Ref. [16]. This technique was successfully applied to obtain the DS curves for rat liver cells in the frequency range from 10$^3$ to 10$^8$ Hz. The authors of this study stated that it remains unclear if the trends in the dispersion curves below 10$^3$ Hz are due to the $\alpha$-effect or to an artifact related to insufficient polarization removal [16]. The method just presented is sometime mentioned under the name of frequency-derivative method.

A more elaborate procedure was proposed in Ref. [14], in which $Z_{\mbox{\tiny{p}}}$ is parameterized as described above and in the same time, $Z_{\mbox{\tiny{s}}}$ is parameterized  using phenomenological frequency dependence functions. The result is a phenomenological fitting function, which is then used to fit the experimental dispersion curves on the entire range of frequencies, using a large number of fitting parameters. This methodology was successfully applied to describe the dielectric properties of erythrocytes in the frequency range from 1kHz to 1GHz. This methodology may be sensitive to the phenomenological function used to parameterize $Z_{\mbox{\tiny{s}}}$, and the results could be biased by a particular choice. For example, the methodology had difficulty in resolving the $\alpha$-dispersion of any live cell suspension.

Distance variation technique is another method for polarization removal. We have used successfully this method in the past to remove the polarization error in low conductivity fluids like pure water, toluene, glycol, and for cells suspended in low conductivity buffers [25]. This methodology is based on the observation that $Z_{\mbox{\tiny{p}}}$ is practically independent of the separation distance $d$ between the capacitor's plates. Thus, by measuring the $Z$=$Z_{\mbox{\tiny{s}}}$+$Z_{\mbox{\tiny{p}}}$ for two values of $d$, one can eliminate $Z_{\mbox{\tiny{p}}}$ by subtracting the results of the two measurements. 

The polarization errors can be reduced by using special electrodes, like the fractal electrodes (platinum black is one example), or using gratings instead of flat electrodes. However, none of these techniques would totally eliminate the polarization error.

As opposed to the above methodologies, our technique give direct, unbiased value of $Z_{\mbox{\tiny{p}}}$. We can use this unbiased measurements to verify the theoretical predictions and the assumptions made in the previous works on the behavior of $Z_{\mbox{\tiny{p}}}$. We find most of them to be true, even at a quantitative level.

\begin{figure*}
 \center
 \includegraphics[width=10cm]{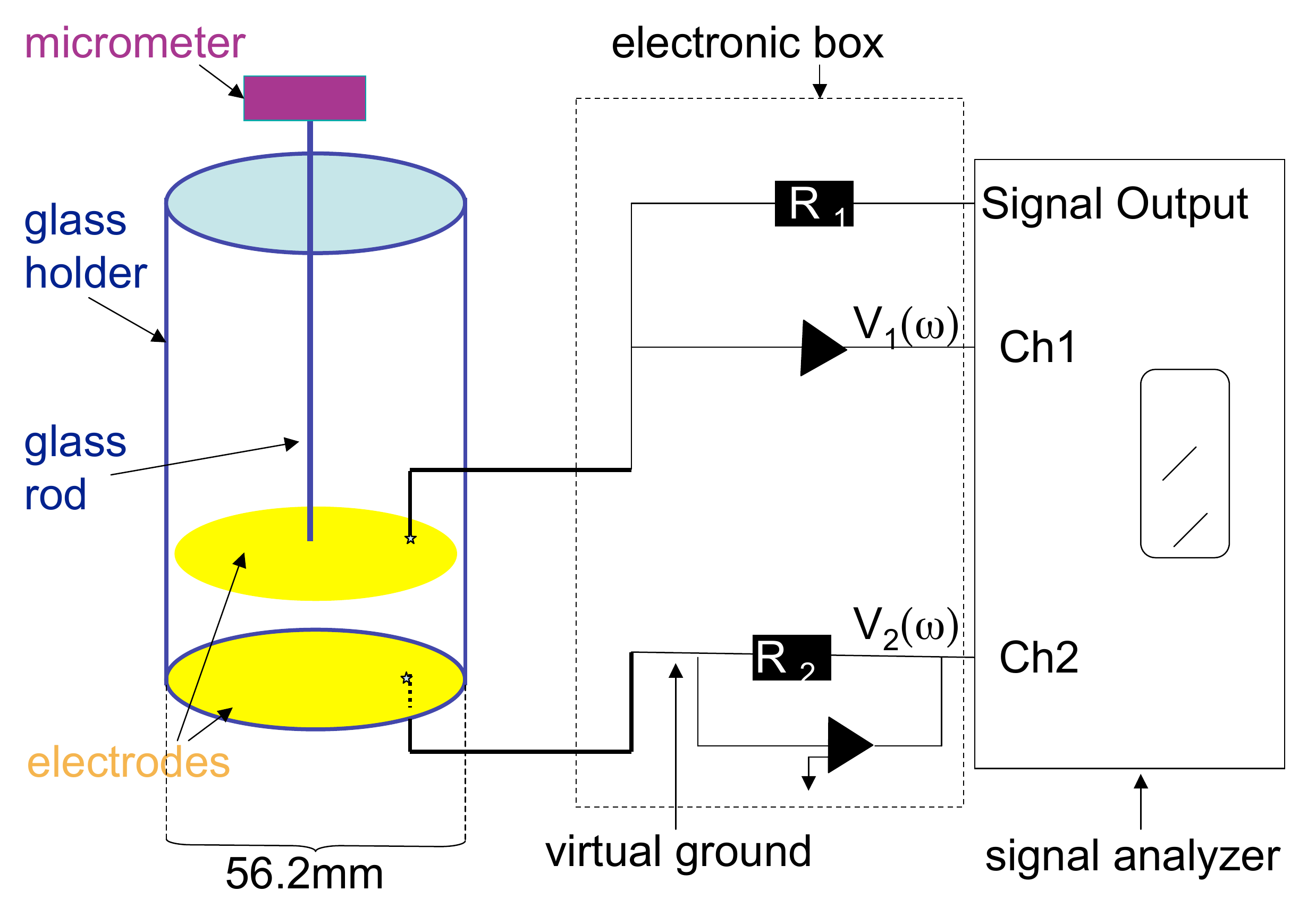}
 \caption{The experimental set up used in this paper. The gold plated electrodes are placed inside a glass cylinder. The distance between them is controlled with a micrometer. A voltage from the signal analyzer is placed on the upper electrode while the bottom one is kept at virtual ground through the negative input of the amplifier. The ration between the voltages at channel 2 and channel one is recorded.}
\end{figure*}

\section{Materials and Method}

\subsection{Experimental Setup}

We use the same experimental setup as in Ref. [25], which is illustrated in Fig. 1. The sample to be measured is placed between two parallel gold plated electrodes (yellow) that are enclosed in a long cylindrical glass tube (dark blue). The results reported here were obtained with electrodes of 28.1 mm in radius. In order to minimize the chemical reactions between the electrodes and the solution, gold, an inert metal, was plated on the electrodes. It was found that 100 nm was enough to withstand the corrosion from the salty buffers and other media used for live cells measurements. The distance between the capacitorÕs plates is controlled with a micrometer. To reduce the errors due to the stray capacitance at the edges of the electrodes, the bottom electrode was made as large as the glass container and the upper electrode just one millimeter smaller, to allow the flow of the fluids when modifying the electrodes separation. Because of this choice, the stray electric field lines will go mostly through the glass container (dielectric permittivity of 4) or air (dielectric permittivity of 1) and, since these are materials or reduced dielectric constant (compared to 78 of water), this reduces the contribution of stray capacitance to the measurements. The electrode guard technique for reducing the stray effects was also tested but no improvement was seen in the measurements. The distance between the electrode plates was varied from 1 mm to 10 mm. 

The experimental setup works as follows. The Signal Analyzer provides a sinusoidal voltage at its signal output. It also digitizes the voltage at Channels 1 and 2 and takes the ratio of these two as function of frequency. The real and imaginary parts of this ratio are stored on a computer. The bottom electrode plate was held at relative ground potential through the negative input of the Amplifier $A_2$. The output voltage from the Signal Analyzer is applied to the upper electrode, through resistor $R_1$. As a result, the current $I$ that flows through the sample produces a voltage $V_1$, equal to the product of $I$ and impedance $Z$ of the measuring cell. The voltage $V_2$ is equal to minus the product of $I$ and resistance $R_2$. Thus the transfer function, $T$, is directly related to the measuring cell impedance by: $T$=$R_2/Z$. $R_2$ was fixed at 100 ½ in these experiments. If the impedance $Z$ were not contaminated by the polarization effects and other possible factors, then the complex dielectric function $\epsilon^*$=$\epsilon$+$\sigma/j\omega$ of the sample could be calculated from:
\begin{equation}
Z=\frac{d/A}{j\omega \epsilon^*}, \ \mbox{or} \ \epsilon^* = \frac{d/A}{j\omega R_2} Z,
\end{equation}
where $d$ represents the distance between the electrodes, $A$ the surface area of one electrode, $\epsilon$ and $\sigma$ are the dielectric permittivity and conductivity of the sample and $\omega$=$2\pi f$, where $f$ is the frequency of the applied signal.

The set-up used in this paper was previously shown to correctly measure samples with a wide range of relative dielectric permittivities, from 2 (toluene) all the way to 78 (water) [25]. Without correcting the the polarization error, the set-up gives the correct value of $\epsilon^*$ for frequencies down to 10 Hz for non-polar liquids and down to 1 kHz for water.

\subsection{E. coli preparation}

In experiments with live cells, E. coli K12 wild type (www.atcc.com) were incubated in 3 ml of Tryptone Soy broth (growth media) at 175 rpm and 37$^o$C to the saturation phase for 16 hours. They were then re-suspended in 60 ml of media and incubated for 5 hours in the same conditions until they reached mid logarithmic phase. To determine the cell growth and concentration, a spectrophotometer was used to measure the optical density of the cell suspension before the measurements. For dielectric spectroscopy, a fresh suspension of cells was centrifuged at 2522$\times$$g$ for 10 min before each measurement and the pellet re-suspended in ultrapure (Millipore) water with 5mM glucose that is routinely used to balance the osmotic pressure.

To check cell viability in media and water, two methods, plate counting and membrane potential dyes, were employed. For plate counting, two test tubes with 10 ml of cells in media and cells in ultrapure water with 5 mM glucose were tested for a extended period of time. Each day, cells were cultivated on agar plates by serial dilutions and incubated for 18 hours at 37$^o$C and the number of colonies counted from the smallest dilution area. The membrane potential dye, Di-SC3-5 Carbocyanine iodide (www.anaspec.com), was used to stain the cell membrane and test for viability. The labeled cells were imaged using an inverted microscope, Axiovert by Zeiss, in the fluorescence mode. Dye was added in a concentration of 0.4 $\mu$M to cells in media and cells in water and imaging was performed after 15 min, to allow the dye to settle in the membrane. All measurements were done at 25$^o$C.

 \begin{figure*}
 \center
 \includegraphics[width=16cm]{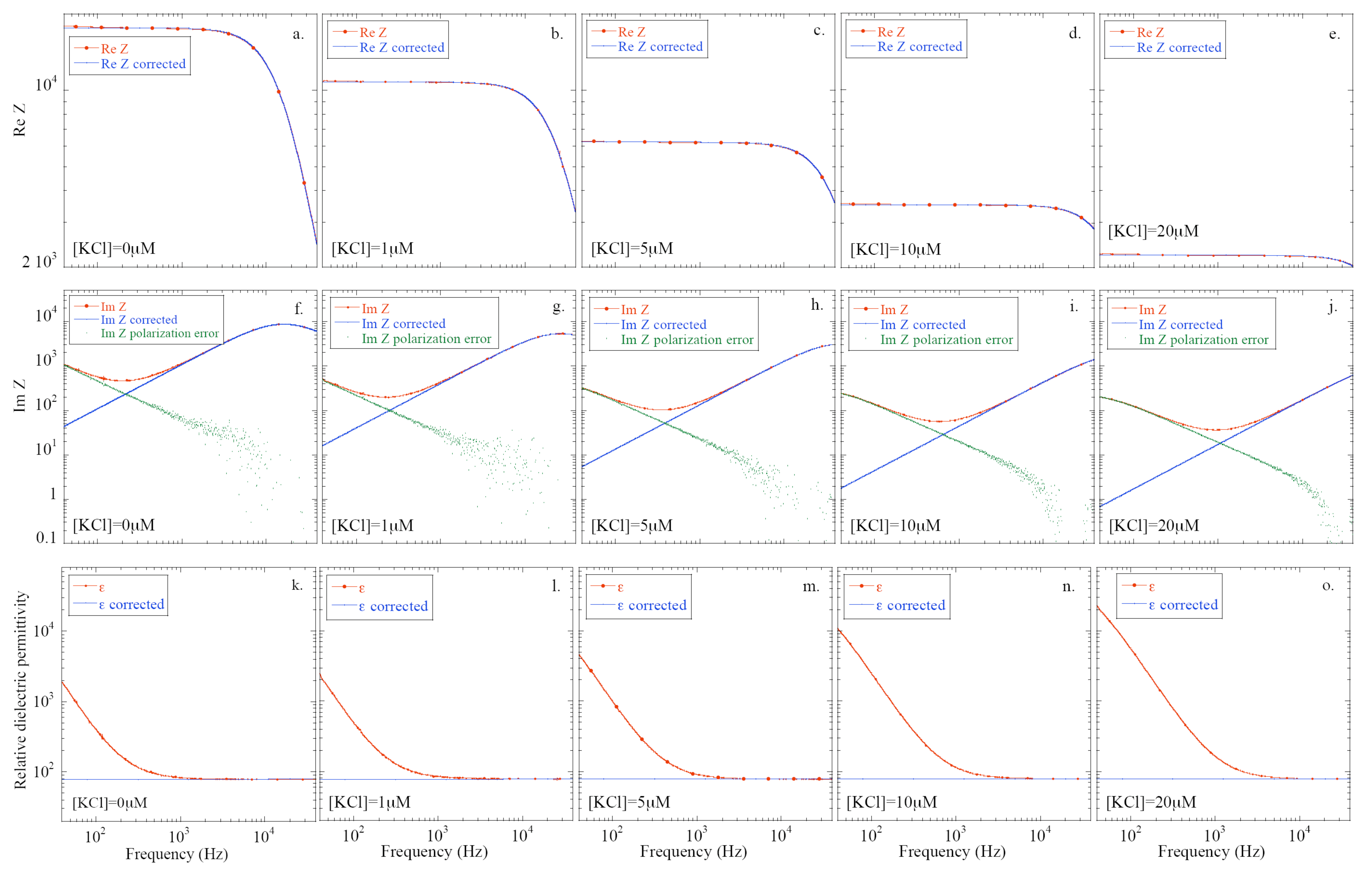}
 \caption{ DS measurements for Milipore water with 0, 1, 5, 10 and 20 $\mu$M KCl. Panels a) - e) show the real part of the impedances; panels f) - j) show the imaginary part of the impedances; panels k) - o) show the corrected and un-corrected dielectric permittivity. Red lines marked with circles represent the dielectric values without compensating for the electrode polarization effects while blue, solid lines represent the dielectric values after removing the polarization effects. The green lines represent the imaginary part of the polarization impedance $Z_{\mbox{\tiny{p}}}$.}
\end{figure*}

\section{Measuring the polarization impedance}

The polarization effect is due to the polarization of the ionic charges accumulated at the interface between the fluid and metallic electrodes. The electrode polarization of the electrical double layer depends on the chemical and physical properties of the electrode as well as of the sample to be measured [26]. Because the effect takes place at the interface between the sample and the electrode, mostly within a few nanometers from the electrode, the polarization impedance $Z_{\mbox{\tiny{p}}}$ appears in series with the intrinsic impedance of the sample $Z_{\mbox{\tiny{s}}}$. In other words, the measured impedance is the sum of the two: $Z$=$Z_{\mbox{\tiny{s}}}$+$Z_{\mbox{\tiny{p}}}$ [26]. Thus, if one develops a method to measure $Z_{\mbox{\tiny{p}}}$, the intrinsic impedance of the sample can be obtained by subtracting $Z_{\mbox{\tiny{p}}}$ from the measured $Z$.

We will use Millipore water to exemplify how one can achieve just that. The dielectric function of pure water is known to be constant $\epsilon_r$=78 (relative to the vacuum) for a frequency range spanning from 0Hz to several GHz. However, if we use the raw DS data and compute the dielectric function using Eq. 1, we obtain the graph shown in Fig. 2k (red line). Looking at this graph, we see that that from approximately 1 kHz and up, $\epsilon_r$ agrees extremely well with the nominal value of 78, while from 1 kHz down, $\epsilon_r$ deviates quite strongly from this value. The anomalous behavior of $\epsilon_r$ at these low frequencies is the typical manifestation of the electrode polarization effects. Now, the main observation is that above 1 kHz the effect is practically gone. If we plot the real and imaginary parts of the impedance $Z$ (the red lines in Fig. 2a and 2f), and the intrinsic impedance of the Millipore water (the blue lines in Figure 2a and 2f), computed as:  $d/j\omega \epsilon^* A)$, with $\epsilon^*$ being the nominal complex dielectric function of Millipore water, we see that the two graphs match quite well above 1 kHz. In fact, if we fit the experimental values of $Z$ above 1 kHz with a frequency dependent function
\begin{equation}
Z_{\mbox{\tiny{fit}}}=\frac{d/A}{j\omega \epsilon +\sigma},
\end{equation}
with $\epsilon$ and $\sigma$ as fitting parameters, we obtain $\epsilon$=78 and $\sigma$=0.000070 S/m, in extremely good agreement with the nominal values of the Millipore water. Now, since it is known that the dielectric function and conductivity of ionic solutions are constant all the way to zero frequency, we can extrapolate $Z_{\mbox{\tiny{fit}}}$ to lower frequencies. The difference between the measured $Z$ and $Z_{\mbox{\tiny{fit}}}$ is precisely the polarization impedance $Z_{\mbox{\tiny{p}}}$. Thus, we can measure in this way the polarization impedance and a plot of $Z_{\mbox{\tiny{p}}}$ is show in the inset of Figure 2f (the green line).

Based on all the above, the following methodology emerges: To measure the polarization impedance for an unknown ionic solution one can do the following:
\begin{enumerate} 
\item Fit the experimental data for $Z$ with the function $Z_{\mbox{\tiny{fit}}}$ given in Eq. 2, by giving a large weight to the high frequency data and a low, or almost zero weight to the low frequency data.
\item From the fit, determine the true dielectric constant $\epsilon$ and conductivity $\sigma$ of the ionic solution.
\item Extrapolate $Z_{\mbox{\tiny{fit}}}$ all the way to 0 Hz and obtain the intrinsic impedance $Z_{\mbox{\tiny{s}}}$ of the ionic solution at low frequencies. 
\item Compute $Z_{\mbox{\tiny{p}}}$ as the difference between $Z$ and the extrapolated $Z_{\mbox{\tiny{s}}}$.
\end{enumerate}
All data on $Z_{\mbox{\tiny{p}}}$ reported in this paper were obtained by an automated implementation of the above four steps. Fig. 2 shows an application of the above procedure to a series of weak saline solutions, which will be analyzed later in the paper.

\section{Characterization of $Z_{\mbox{\tiny{p}}}$} 

We will use the Millipore water to answer several important questions about the polarization impedance. As we already mentioned, $Z_{\mbox{\tiny{p}}}$ depends, in general, on the applied voltage and on frequency. Fig. 3 maps the dependence of $Z_{\mbox{\tiny{fit}}}$ for Millipore water on these parameters. The methodology outlined above was applied for three values of the distance between capacitor plates, $d$=1, 3 and 5 mm. For each $d$, the experiments were repeated for five values of applied voltages per centimeter: 0.1, 0.3, 0.5, 0.7 and 0.9 V/cm. The steps i)-iv) listed above were applied to the resulting 30 measurements. 

 \begin{figure*}
 \center
 \includegraphics[width=13cm]{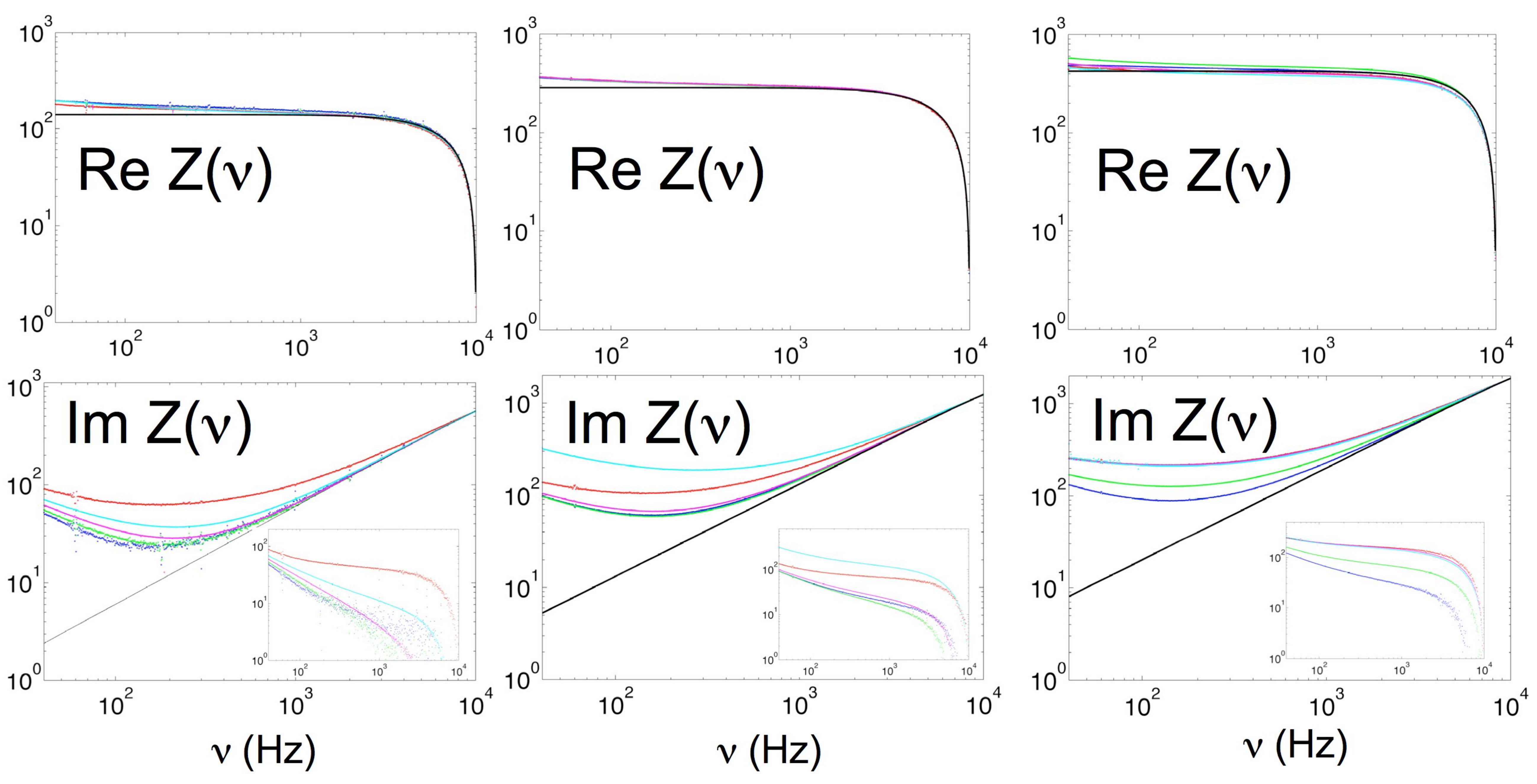}
 \caption{ Real (1st row) and imaginary (2nd row) parts of the total impedance for d=1mm (1st column), d=3mm (2nd column) and d=5mm (3rd column), for applied electric fields of 0.1 (red), 0.3 (blue), 0.5 (green), 0.7 (magenta) and 0.9 (cayan) V/cm. Black line marks Zideal. The insets represent Im Zp for the above distances and electric fields.}
\end{figure*}

From the first row of Fig. 3, one can see Re[$Z_{\mbox{\tiny{fit}}}$] almost overlapping with Re[$Z$], while on the second row one can see a large difference between Im[$Z_{\mbox{\tiny{fit}}}$] and Im[$Z$]. Thus, our first finding is that $Z_{\mbox{\tiny{p}}}$ is mostly reactive. This confirms the assumption on $Z_{\mbox{\tiny{p}}}$ made on Ref. 14 that the coefficient $A$ can be set to zero. The insets in Fig. 3 show Im[$Z_{\mbox{\tiny{p}}}$] as function of frequency in a log-log scale. The plots are for various values of $d$ and applied electric fields. The Im[$Z_{\mbox{\tiny{fit}}}$] was obtained from the main plots by taking the difference Im[$Z_{\mbox{\tiny{fit}}}$]=Im[$Z$]-Im[$Z_{\mbox{\tiny{fit}}}$]. The log-log plots appears as straight lines for a wide range of frequencies, fact that confirms the power law behavior of Im[$Z_{\mbox{\tiny{fit}}}$] with the frequency. The exponent of the power law can be easily and accurately obtained from the plots and will be disccused later in the paper.
Analyzing the insets, we observe a weak dependence of Im[$Z_{\mbox{\tiny{fit}}}$] on the applied electric field. The shape of the curves changes slightly as the electric field is varied and this effect seems to be more pronounced for the smallest value of $d$. For larger $d$, the effect is almost nonexistent which supports the assumption that the amplitude and the exponent of the power law are independent of the applied electric field. 

\section{Robustness of the methodology}
 
\subsection{Weak ionic solutions}

In Fig.~2 we present an application of the methodology to electrolytes with low conductivity, more precisely, water with low concentrations of KCl. From left to right, the different columns in Figure 3 refer to 0 $\mu$M, 1 $\mu$M, 5 $\mu$M, 10 $\mu$M and 20 $\mu$M KCl. For low KCl concentration, the dielectric function of the solution remains constant at 78, while large increases in the conductivity should be observed. In all columns of Fig.~2, we can see an almost perfect match between the real parts of $Z$ and $Z_{\mbox{\tiny{fit}}}$. The fit is also perfect for the imaginary parts of $Z$ and $Z_{\mbox{\tiny{fit}}}$, if we look above 1 kHz. The fitting provided the following values: $\epsilon_r$=78$\pm$1,  and  $\sigma$=0.00011, 0.00020, 0.00034 and 0.00056 S/m for the four cases, respectively. {\it To these significant digits, same values of $\sigma$ have been measured using a regular conductivity probe} (Denver Instruments Model 220). This demonstrates that, by an automated implementation of the steps i)-iv) described above, we were able to remove the polarization effects and obtain the intrinsic dielectric function and conductivity of all these solutions. 

\subsection{Strong ionic solutions}

Low frequency dielectric spectroscopy has applications in many fields, but we are primarily interested in biological applications, such as measuring the dielectric functions of live cell suspensions. Many physiological buffers are known to have huge conductivities. Thus, in order to apply our method to live cells in suspension, we must demonstrate that it works for strong ionic solutions.
 
As the conductivity of the electrolyte increases so does the frequency limit where the polarization effect highly contaminates the data. Since our method relies on the information contained in the high frequency domain where the polarization effects are small, the experimental data must contain a good part of this domain. Thus, for strong ionic solutions, we need to record the DS curves up to much higher frequencies. This is why, in the experiments with strong ionic solutions we replaced the initial signal analyzer SR 795 with the Solartron 1260, which allowed us to record data in the frequency window from 0 Hz to 1 MHz. 

We give an application of the methodology to water and KCl, at much higher concentrations than before, namely 0.1, 0.5 mM KCl concentrations, and to HEPES with 5 mM concentration. The dielectric functions before and after polarization removal are shown in the left panel of Fig. 4. For the two KCl solutions, the methodology provides the following values:  $\epsilon_r$=72$\pm$1 and   $\sigma$=0.00557 and 0.00707 S/m. For HEPES solutions, the methodology provides $\epsilon_r$=71  and  $\sigma$=0.00482 S/m. Again, {\it to these significant digits, the same values of $\sigma$ have been measured using the conductivity probe}. 

According to our measurements, the dielectric constant of the KCl solution decreases with the KCl concentration, an effect that has been reported before. This has to do with the fact that K+ and Cl- ions donÕt have an intrinsic dipole moment (meaning a smaller polarizability than water molecules) and since they occupy space in the solution the overall dielectric permittivity of the solution decreases with the KCl concetration [27].

\subsection{Behavior of $Z_{\mbox{\tiny{p}}}$ with the ionic concetration}
 
Now we can say a few things about the behavior of  $Z_{\mbox{\tiny{p}}}$ with the increase of KCl concentration. In Fig.~5 we put together the Im[$Z_{\mbox{\tiny{p}}}$] reported in Figs.~2 and 3. Looking at Fig.~5, one can see two manifolds of curves corresponding to low and high KCl concentrations. The manifold of low KCl concentrations is higher and more spread out when compare to the manifold for higher KCl concentrations. All curves show a consistent decrease of Im[$Z_{\mbox{\tiny{p}}}$] with the increase of KCl concentration. Quantitative analysis of this decrease show that Im[$Z_{\mbox{\tiny{p}}}$] is {\it inversely proportional to the KCl concentration}, in line with theoretical studies that also found Im[$Z_{\mbox{\tiny{p}}}$] to be inversely proportional with the condu ctivity of the electrolyte [14]. This explains why the manifold of curves for low KCl concentrations is more spread out.

Another finding is that the exponent $\alpha$ of the power law behavior Im[$Z] \propto \omega^{-\alpha}$, {\it decreases} with the increase of KCl concentrations. This decrease is visible even within each manifold. Quantitatively, we find an average of $\alpha$=0.8 for low KCl and an average of $\alpha$=0.6 for higher KCl concentrations. The last value is in line with a previous study [14], which found an exponent of $\alpha$=0.7.14 for mM KCl concentrations.  Fig.~5 also presents Im[$Z_{\mbox{\tiny{p}}}$] for a HEPES buffer. Although the solutions of HEPES and mM KCl have similar conductivities, the imaginary parts of $Z_{\mbox{\tiny{p}}}$ for HEPES behaves slightly different. For the HEPES concentrations used here, Im[$Z_{\mbox{\tiny{p}}}$] has a power law behavior with an exponent of 0.5 only after 400 Hz. Below this frequency the power law breaks down. HEPES are zwitterions, electrically neutral molecules but with negative and positive charges on different atoms of the same molecule as opposed to KCl which form monovalent ions in solution. This difference may lead to the different behavior of the polarization impedance seen in our experiments.

 \begin{figure}
 \center
 \includegraphics[width=8cm]{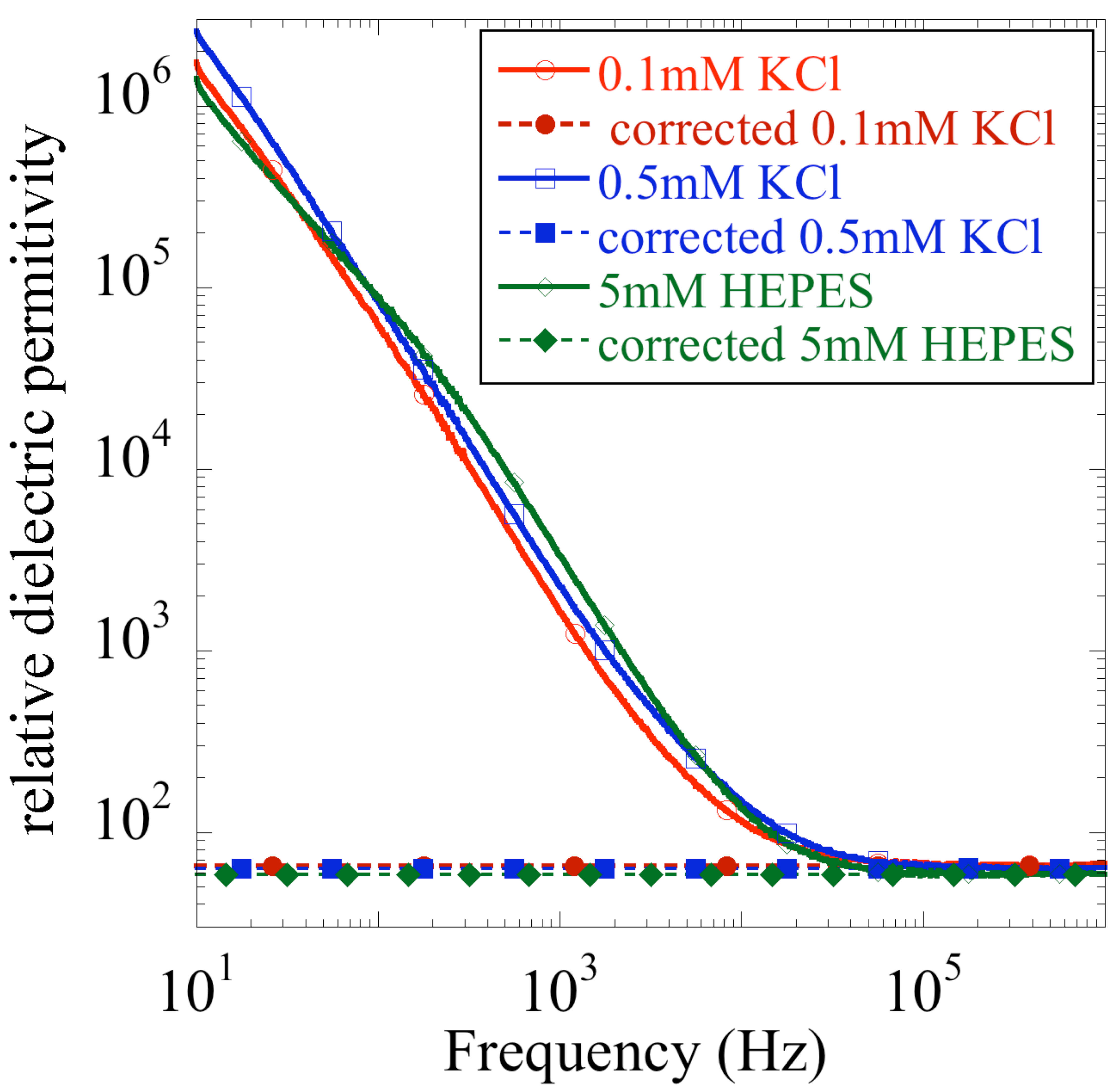}
 \caption{Dielectric permittivity versus frequency for (left) water with 0.1 and 0.5 mM KCl and 5mM HEPES and (right) E.coli cell suspension for an optical density (O.D.) of 0.051. The dotted lines represent the real relative dielectric permittivity after the removal of the polarization error.}
\end{figure}

\section{Polarization effects in colloidal suspensions}

We should point out from the beginning that the dielectric functions of colloidal suspensions are not constant with frequency. In fact, the main goal of dielectric spectroscopy is to capture and study these variations of ? with frequency. Thus, the methodology outlined above cannot be directly applied to these systems. However, one way to apply the methodology to colloids (including live cells) suspended in electrolytes or to samples saturated with electrolytes is as follows. Initially, the sample is measured and the raw values of Z are recorded. The data are, of course, contaminated by the polarization effect. Now the key observations are: 1) {\it the polarization effect is due to the electrolyte and not the colloidal particles} (as discussed below) and 2) {\it the dielectric function of the electrolyte is constant with frequency}. Thus, if we separate the electrolyte from the colloids, we can measure $Z_{\mbox{\tiny{p}}}$ as before, which then can be removed from $Z$ to obtain the intrinsic impedance of the sample. Thus, the following methodology emerges:
\renewcommand{\theenumi}{\Roman{enumi}}
\begin{enumerate}
\item Record the raw values of $Z$.
\item Gently remove the colloids from the solution and save the supernatant.
\item Apply steps i)-iv) to the supernatant and determine $Z_{\mbox{\tiny{p}}}$.
\item Remove $Z_{\mbox{\tiny{p}}}$ from $Z$ to obtain the intrinsic impedance of the sample.
\end{enumerate}
An important question arises, namely, if the polarization impedance seen in the supernatant is the same as the one in the cell suspension. We argue that, for the suspensions of cells used in our experiments, the polarization error is due to the supernatant. Indeed, the polarization impedance is due to the polarization of the ionic double layer near the electrodes when time oscillating electric fields are applied. The thickness of electrical double layer responsible for the polarization error is a few nanometers from the electrode. In our experiments, the applied electric field is kept small, below 1V/cm, so it does not rotate or move the cells. Consequently, the cells do not concentrate on the electrodes when the voltage is turned on. Since the usual cell concentrations are relatively small, the average number of cells in the electrical double layer is practically zero. The only logical conclusion is that the cells do not contribute to the polarization error. 

It is also important to notice how we propose to measure the dielectric properties of the electrolyte in which the cells are suspended. When the cells are suspended in an electrolyte, they modify the dielectric properties of the electrolyte by releasing ions through an ion-exchange process needed to stabilize their membrane potential. Thus, the properties of the electrolyte change when the cells are suspended and this is why we must measure the electrolyte after the cells have been suspended. 

Another important observation is that the protocol I-IV can be applied in one step. For example, for mamalian cells that are large in size, it is possible to divide the measuring cell in two compartments using a membrane that is permeable to the electrolyte but not to the live cells. Such a measuring cell, illustrate in Fig.~6, will allow the separation of the electrolyte and simultaneous DS measurements for the cell suspension and for the electrolyte. The measuring cell of Fig.~6 is easy to construct for geophysical applications where one is interested in the dielectric properties of granular soils, with granules typically of millimeter or slightly sub-millimeter sizes. We can argue that such a measuring cell will be superior to the distance variation cells because there are no moving parts and because we obtain the value of $Z_{\mbox{\tiny{p}}}$, which is an interesting quantity in itself. 

 \begin{figure}
 \center
 \includegraphics[width=8cm]{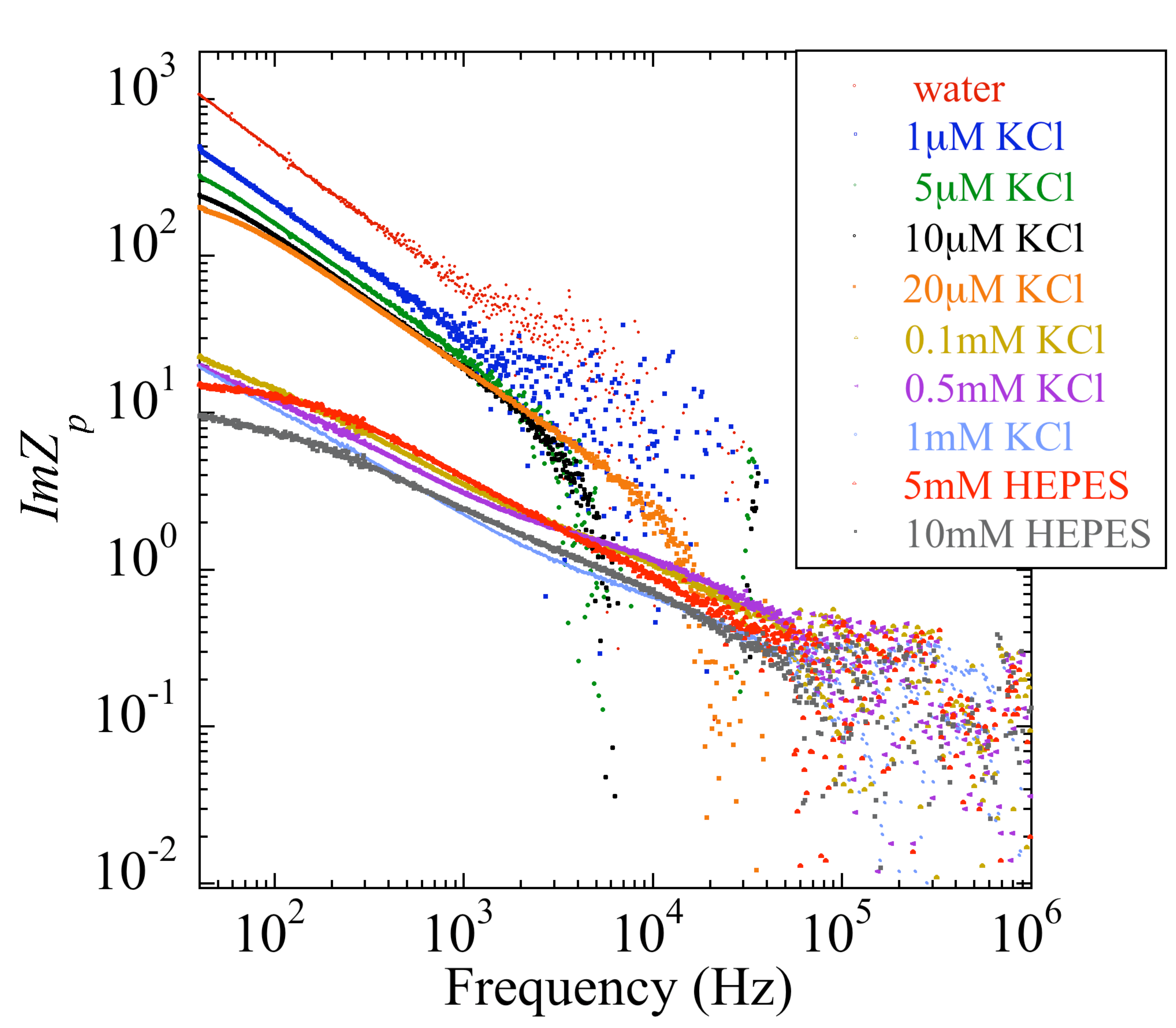}
 \caption{Imaginary part of the electrode polarization impedance for the samples from Figures 3 and 6 as function of frequency. The curves show a power law behavior with an average exponent equal to 0.8 for ?M KCl concentrations, 0.6 for mM KCl concentrations and 0.5 for HEPES. The amplitude goes inversely proportional with the conductivity, as discussed in text.}
\end{figure}

Since the mamalian cells are more tedious and expensive to grow, we show here an application of the above protocol to live E. coli cell suspensions. E. coli cells are 2 to 5 $\mu$m in size so the measuring cell illustrated in Fig.~6 is not feasible. Instead, we chose to implement the protocol I-V in two steps. For this, we first take the DS measurements on the E. coli suspension and then separate the supernatant by gentle centrifugation and subject it to the DS measurements. During the experiments, we observed that if the cells are not carefully spun, the ionic properties of the supernatant can be different from that of the cell solution. In particular, the conductivity of the supernatant becomes higher than that of the cell solution, suggesting that ions were released in the electrolyte during the centrifugation. 

For this reason, we must demonstrate that we can successfully separate the supernatant without changing its ionic properties. This can be done because we have a control parameter, the conductivity, which must be the same, with a high degree of accuracy, for the supernatant and for the cell suspension. This is exactly what is observed when the the centrifugation speed is reduced, namely that the conductivity of the supernatant comes closer and closer to that of the cell suspension. This is our experimental evidence that the supernatant has the same ionic properties as that of the cell suspension, and consequently, the polarization of the double layer near the electrodes should be similar for the two cases. We also provide the error bars, from which one can determine the fluctuations induced by the two step implementation. We mention that the error bars are small enough to allow quantitative evaluations of the membrane potential during various induced physiological changes in the E. coli cells [x]. 

  \begin{figure}
 \center
 \includegraphics[width=8cm]{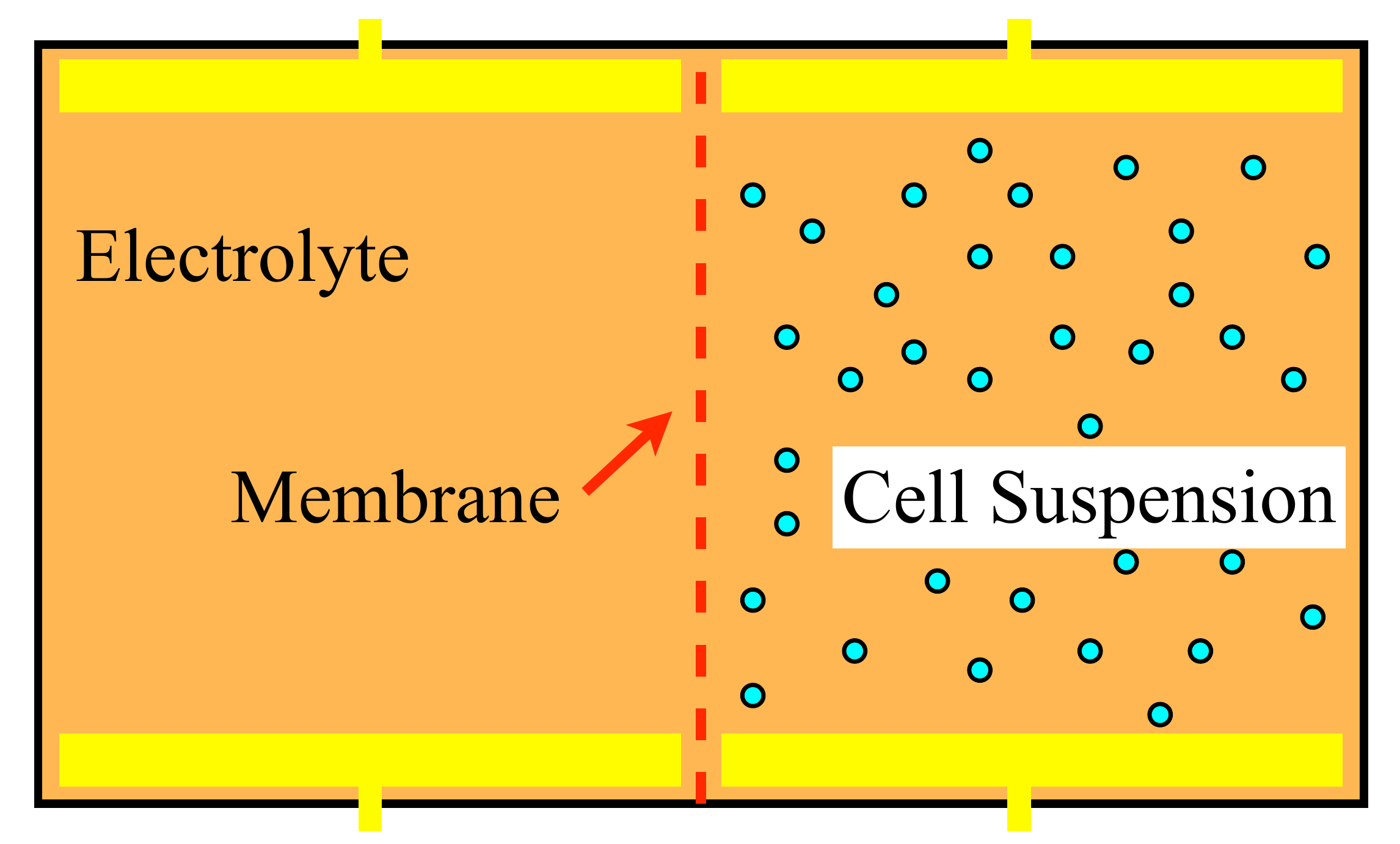}
 \caption{A measuring cell that will allow simultaneous DS measurements of the live cells suspension (using the right capacitor) and electrolyte (using the left capacitor).}
\end{figure}

The E-coli cells were harvested in their mid logarithmic phase then re-suspended for measurements in ultra pure water plus 5mM glucose (for osmotic pressure balancing) at an OD=0.186. The sample was prepared as explained in Section 2. Same section also describes our measurements done to test the cell suspension viability, which proved that the cellular suspension was healthy. The measurements were taken with a distance between the electrode plates of 3 mm. In Fig. 7a, we present the raw values (magenta) of the imaginary part of the measured impedance $Z$ and that of the impedance of the supernatant (red) obtained by extremely gentle centrifugation. The inset gives a detailed picture of the data in the frequency range from 20 to 100 Hz, plotted in a lin-lin scale. The plot shows a substantial difference between the two sets of data (of more than 5 \%), difference that can be easily resolved by our electronic setup. The green line in Fig.~8a represents the imaginary part of $Z_{\mbox{\tiny{p}}}$ determined from the DS measurements on the supernatant as explained in the I-IV protocol. The blue line represents the imaginary part of the intrinsic impedance $Z_{\mbox{\tiny{s}}}$ of the cell suspension.

Fig.~7b shows the dielectric permittivity of the E. coli suspension as computed from $Z$ (magenta) and computed from $Z_{\mbox{\tiny{p}}}$ (blue). The blue line represents the intrinsic DS curve of the sample. Fig.~8 shows the intrinsic DS curve of two E. coli solutions of different volume concentrations, each obtained by averaging 5 independent measurements on 5 separate samples. The figure also shows the error bars resulted from the 5 measurements. As one can see, the error bars (less than 5 \%) are small enough to allow the resolution of the two concentrations. An more complete analysis, including comparisons with the existing theoretical models [9,11] is given in Ref.~[28]. In this reference we show that the experimental DS curves are in line with the theory and that we can obtain quantitative values for the membrane potential of the E. coli cells in suspension.

Examining Fig.~7 and 8, we can distinguish a very high plateau in the DS curve, which we identify with the $\alpha$-plateau [9,11]. An  intriguing feature of the DS curves is the absence of the $\beta$-plateau. To explain the absence of the $\beta$-plateau, we mention that the dielectric dispersion curves of cell suspensions are very sensitive to the conductivity of the medium. In particular, the $\beta$-dispersion shifts to the lower frequencies as this conductivity is decreased [11]. In our measurements, the cells were suspended in pure water, thus the $\beta$-dispersion should be substantially shifted to the left. The $\beta$ dispersion of E. coli solutions was measured in Ref.~[29] and found to be between $10^5-10^6$ Hz. In these experiments, the cells were suspended in growth medium with conductivity of about 440 times larger than that of the water used in our experiments. Using a well established model of the $\beta$ response, we found that when lowering the conductivity by 440 times, the $\beta$ dispersion moves from $10^5-10^6$ Hz in the region where we observe the $\alpha$ plateau [28]. Thus, in our experiments, the $\beta$ dispersion is covered by the $\alpha$ dispersion due to the low conductivity of the medium in which the cells were suspended.

\begin{figure}
 \center
 \includegraphics[width=10cm]{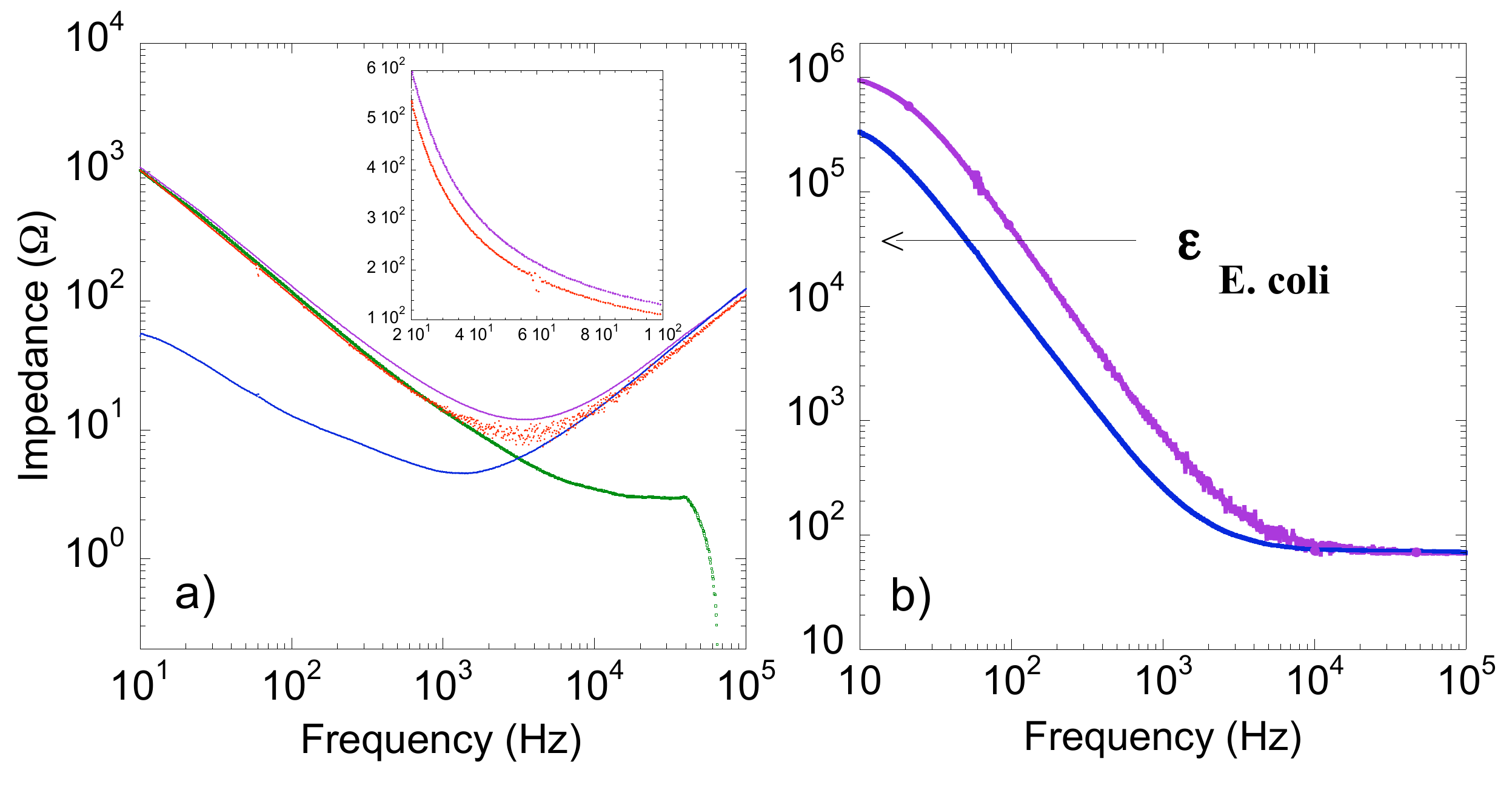}
 \caption{Dispersion curves for E. coli at OD=0.186. Left: the imaginary part of the measured impedance $Z$ of the cell suspension (mangenta), of the supernatant (red), of the polarization impedance $Z_{\mbox{\tiny{p}}}$ (green) and of the corrected (intrinsic)impedance $Z_{\mbox{\tiny{p}}}$ of the sample (blue). Right: The dielectric curves for the cell
suspension before (mangenta) and after the correction for the electrode polarization impedance
(blue) done using the technique presented in this paper.}
\end{figure}

\section{Conclusions}

In this paper we proposed a methodology to measure, analyze and remove the polarization impedance $Z_{\mbox{\tiny{p}}}$ that appears and contaminates the dielectric spectroscopy data at low frequencies. Using this methodology we mapped the dependence of $Z_{\mbox{\tiny{p}}}$ on different parameters. Our study experimentally confirmed the theoretical predictions that the polarization impedance is reactive and that it varies as a power law with frequency for monovalent ionic solutions. Our study also showed that the amplitude and the exponent of the power law are very weakly dependent on the applied voltage. We found that the exponent of the power law and the amplitude decreases with the increase of ionic concentration in quantitative agreement with previous works. By comparing two different types of solutions that have similar conductivities, KCl and HEPES we saw that the polarization impedance can have slightly different behaviors. Thus, we pointed out that one must be careful with the assumptions made on the polarization impedance. 

We propose the protocol I-IV that will allow the application of the methodology to colloidal suspensions. We argued that the protocol can be implemented in one step by using the measuring cell shown in Fig.~6.  Such implementation is currently under the investigation. Here, we have demonstrated a two step implementation, in which the cells are removed by gentle centrifugation and the resulting supernatant is used to measure $Z_{\mbox{\tiny{p}}}$. We have presented evidence that, by gentle centrifugation, the resulting supernatant retains the same ionic properties as the cell suspension. We also presented evidence that the two step implementation is robuts, leading to small error bars even when repeating the measurements on completely new samples.   

Obtaining uncontaminated dielectric spectroscopy data in the $\alpha$-region opens the possibility of many interesting applications, the most notable being measuring the membrane potential as predicted in Ref.9.

\begin{figure}
 \center
 \includegraphics[width=8cm]{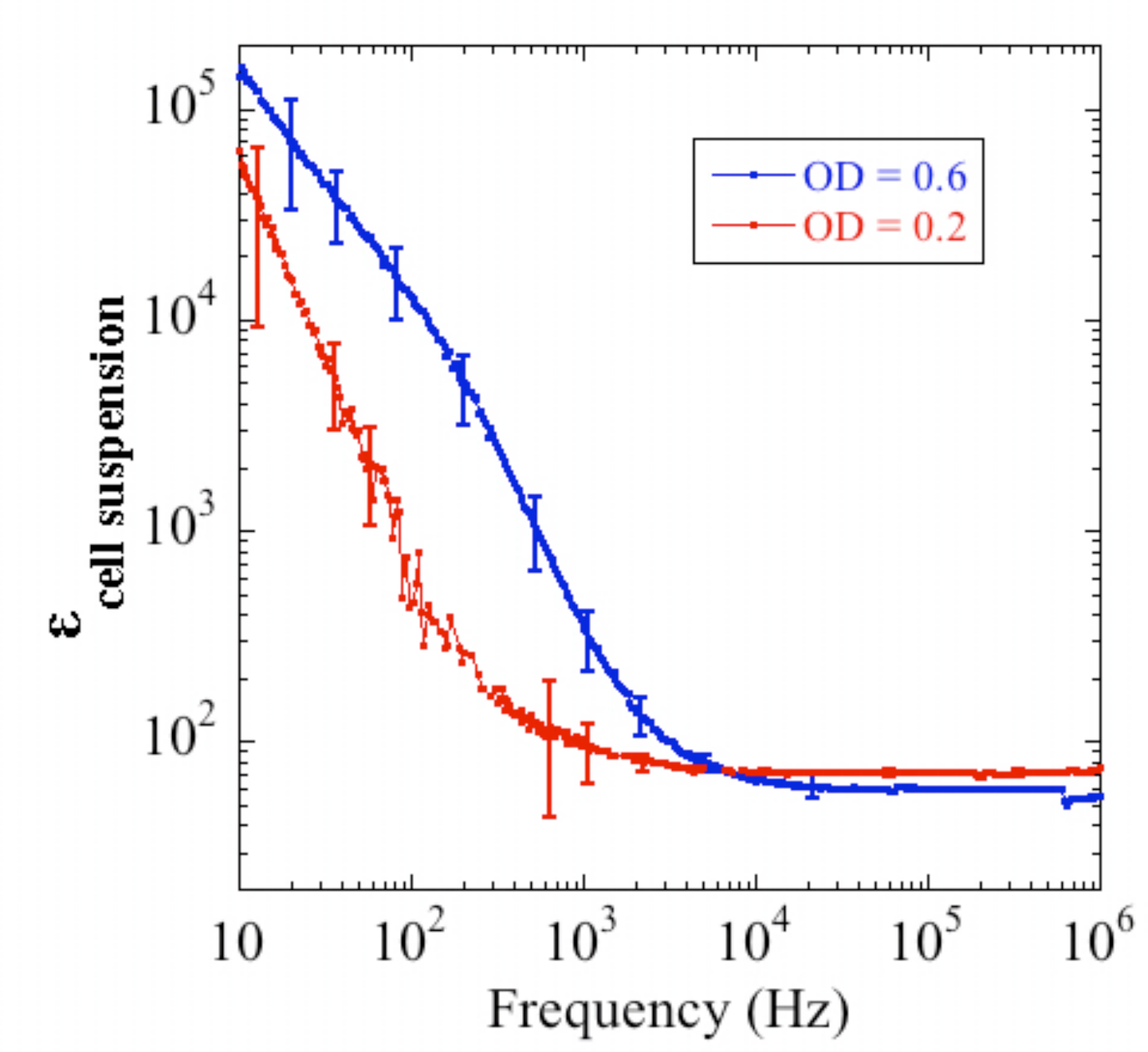}
 \caption{The dielectric curves for two cell suspensions of different volume concentrations. The error bars were obtained by repeating the DS measurements on five fresh samples.}
\end{figure}

\noindent {\bf Acknowledgement.} This work was supported by a grant from NJIT-ADVANCE that is funded by the National Science Foundation (grant Nr. 0547427).

\bigskip
 
\noindent {\bf References.}

\noindent 1	G. Leung, H. R. Tang, R. McGuinness, et al., Journal of the Association for Laboratory Automation, 258 (2005).

\noindent 2	D. Nawarathna, J. R. Claycomb, J. H. Miller, et al., Applied Physics Letters, 86, 023902 (2005).

\noindent 3	G. R. Facer, D. A. Notterman, and L. L. Sohn, Applied Physics Letters, 78, 996 (2001).

\noindent 4	D. Nawarathna, J. H. Miller, and J. Claycomb, Physical Review Letters, 95, 158103 (2005).

\noindent 5	C. R. Keese, J. Wegener, S. R. Walker, et al., Proceedings of the National Academy of Sciences, 101, 1554 (2004).

\noindent 6	L. L. Sohn, O. A. Saleh, G. R. Facer, et al., Proceedings of the National Academy of Sciences of the United States of America, 97, 10687 (2000).

\noindent 7	L. Slater, D. Ntarlagiannis, Y. Personna, et al., Geophysical Research Letters, 34 (2007).

\noindent 8	L. Slater, D. Ntarlagiannis, and D. Wishart, Geophysics, 71, A1 (2006).

\noindent 9	C. Prodan and E. Prodan, Journal of Physics: D, 32, 335 (1999).

\noindent 10	C. Grozze and V. Shilov, J. of Colloids and Interface Science, 309, 283 (2007).

\noindent 11	E. Prodan, C. Prodan, and J. H. Miller, Biophysical J., 95, 1 (2008).

\noindent 12	J. Gimsa and D. Wachner, Biophysical J., 75, 1107 (1998).

\noindent 13	F. Carrique, F. Arroyo, V. Shilov, et al., J. of Chemical Physics 126 (2007).

\noindent 14	F. Bordi, C. Cametti, and T. Gili, Bioelectrochemistry, 54, 53 (2001).

\noindent 15	C. L. Davey and D. Kell, Bioelectrochemistry. Bioenergetics, 46, 91 (1998).

\noindent 16	V. Raicu, T. Saibara, and A. Irimajiri, Bioelectrochemistry. Bioenergetics, 48, 325 (1998).

\noindent 17	H. P. Schwan and C. D. Ferris, Review of Scientifical Instruments, 39, 481 (1968).

\noindent  18	C. L. Davey, G. H. Markx, and D. B. Kell, European Biophysical J., 18, 255 (1990).

\noindent 19	H. Sanabria and J. H. Miller, Physical Review E , 74 (2006).

\noindent 20	M. Umino, N. Oda, and Y. Yasuhara, Medical and Biological Engineering and Computing 40, 533 (2002).

\noindent 21	M. R. Stoneman, M. Kosempa, W. D. Gregory, et al., Physics in medicine and biology, 52, 6589 (2007).

\noindent 22	U. Kaatze and Y. Feldman, Measurement Science and Technology, 17, R17 (2006).

\noindent 23	Y. Feldman, E. Polygalov, I. Ermolina, et al., Meas. Sci. Technol., 12, 1355 (2001).

\noindent 24	H. P. Schwan, Ann. NY Acad of Sci, 148, 191 (1968).

\noindent 25	C. Prodan, F. Mayo, J. R. Claycomb, et al., Journal of Applied Physics, 95, 3754 (2004).

\noindent 26	U. Kaatze and Y. Feldman, Meas. Sci. Technol, 17, R17 (2006).

\noindent 27	W. R. Fawcett, Liquids, Solutions, and Interfaces (Oxford University Press, New York, 2004).

\noindent 28      Corina Bot and Camelia Prodan, arXiv:0812.3352 (2008).

\noindent 29 Bai, W., K.S. Zhao, and K. Asami, Biophysical Chemistry, 2006. {\bf 122}: p. 136-142.

\end{document}